\documentclass[12pt]{spieman}  
\usepackage{amsmath,amsfonts,amssymb}
\usepackage{graphicx}
\usepackage{setspace}
\usepackage{tocloft}
\usepackage[outdir=./]{epstopdf}
\usepackage{subcaption}
\usepackage{wrapfig}
\usepackage{lineno}
\usepackage{aas_macros}
\usepackage{multirow}
\usepackage[dvipsnames]{xcolor}
\usepackage{cancel}
\graphicspath{{./}}
\DeclareGraphicsExtensions{.eps}

\title{Experimental verification of off-axis polarimetry with Cadmium Zinc 
Telluride detectors of AstroSat-CZT Imager}

\author[a,*]{C. S. Vaishnava}
\author[a]{N. P. S. Mithun}
\author[a,$\dagger$]{Santosh V. Vadawale}
\author[a,b]{Esakkiappan Aarthy}
\author[a]{Arpit R. Patel} 
\author[a]{Hiteshkumar L. Adalja}
\author[a]{Neeraj Kumar Tiwari}
\author[a]{Tinkal Ladiya}
\author[c,d]{Nilam Navale}
\author[e]{Tanmoy Chattopadhyay}
\author[f,g]{A. R. Rao}
\author[h]{Varun Bhalerao}
\author[g] {Dipankar Bhattacharya}

\affil[a]{Physical Research Laboratory, Ahmedabad, Gujarat, India, 380009}
\affil[b]{Indian Institute of Technology, Gandhinagar, Gujarat, India, 382355}
\affil[c]{University Department of Physics, University of Mumbai, Maharashtra, India, 400098} 
\affil[d]{Ramniranjan Jhunjhunwala College, Ghatkopar, Mumbai, India, 400086}
\affil[e]{Kavli Institute of Particle Astrophysics and Cosmology, Stanford University, California, United States, 94305} 
\affil[f]{Tata Institute of Fundamental Research, Mumbai, Maharashtra, India,400005}
\affil[g]{Inter University Centre for Astronomy and Astrophysics, Pune, Maharashtra, India, 411007}
\affil[h]{Indian Institute of Technology, Bombay, Maharashtra, India, 400076}

\cftpagenumbersoff{figure}
\cftpagenumbersoff{table}  

\newcounter{daggerfootnote}
\newcommand*{\daggerfootnote}[1]{%
    \setcounter{daggerfootnote}{\value{footnote}}%
    \renewcommand*{\thefootnote}{\fnsymbol{footnote}}%
    \footnote[2]{#1}%
    \setcounter{footnote}{\value{daggerfootnote}}%
    \renewcommand*{\thefootnote}{\arabic{footnote}}%
    }

\begin{document} 
\maketitle
 
\begin{abstract}  
 The Cadmium Zinc Telluride Imager (CZTI) onboard  AstroSat consists of an array of a large number of
 pixellated CZT detectors capable of measuring the polarization of incident hard X-rays. The polarization measurement 
 capability of CZTI for on-axis sources was experimentally confirmed before the launch. CZTI has yielded tantalizing
 results on the X-ray polarization of the Crab nebula and pulsar in the energy range of 100 -- 380 keV. CZTI has also
 contributed to the measurement of prompt emission polarization for several Gamma-Ray Bursts (GRBs). However,
 polarization measurements of off-axis sources like GRBs are challenging. It is vital to experimentally calibrate the
 CZTI sensitivity to off-axis sources to enhance the credence of the measurements. In this context, we report the verification
 of the off-axis polarimetric capability of pixellated CZT detectors through the controlled experiments carried
 out with a CZT detector similar to that used in CZTI and extensive Geant4 simulations of the experimental set-up. Our
	current results show that the CZT detectors can be used to measure the polarization of bright GRBs up to off-axis angles of $\sim$60$^\circ$. However, at incidence angles between 45-60$^{\circ}$, there might be some systematic effects which needs to be taken into account while interpreting the measured polarisation fraction.

\end{abstract} 

\keywords{X-ray polarimetry, off-axis polarization, CZT detectors, GRB} 

{\noindent \footnotesize\textbf{*} SreeVaishnava Cherukuri,  \linkable{vaishnav@prl.res.in} , \linkable{sreevaishnava1997@gmail.com} }\\
{\noindent \footnotesize\textbf{$\dagger$} Santosh Vadawale,  \linkable{santoshv@prl.res.in} }

\begin{spacing}{2}   

\section{Introduction}
\label{sect:intro}  
X-ray astronomy has grown tremendously over the past five decades, with imaging, spectroscopy,
and timing sensitivities comparable to those at other wavelengths. However, the field of X-ray
polarimetry is still largely unexplored due to the inherent complexities in measuring X-ray polarization \cite {weisskopf18}. 
Despite the scientific importance of X-ray polarimetry \cite {krawczynski11}, there have been very few dedicated X-ray polarimeters since the first reliable measurement of the X-ray polarization of the Crab nebula at 2.6 and 5.2 keV by OSO-8 \cite{weisskopf78}. 
Several attempts were made to use the polarization capability of spectroscopic instruments such as RHESSI, IBIS and SPI onboard INTEGRAL, in addition to dedicated balloon borne polarimeters like X-Calibur \cite {guo13}, POGOlite \cite {kamae08} and POGO+ \cite {friis18}.
Measurements of bright sources, Crab \cite {forot08} and Cygnus X-1 \cite {jourdain12}, have been reported with these instruments.
One of the reasons for the slow progress in X-ray polarimetry is that X-ray polarization 
measurements are highly prone to systematics and require a large number of photons. However, the field of X-ray polarimetry 
for persistent X-ray sources is expected to get an impetus 
in the near future with dedicated missions such as the recently launched IXPE~\cite {weisskopf16,weisskopf22} and upcoming XPoSat~\cite {vadawale10,maitra11}~\protect\daggerfootnote{\url{http://www.rri.res.in/~bpaul/polix.html}}.
IXPE is expected to provide two orders of 
magnitude improvement in sensitivity within the energy range of 2 -- 8 keV 
\cite {stephen18} over the earlier OSO-8 measurements, 
whereas the XpoSat mission will extend the energy range to 8 -- 30 keV.

The polarization measurements for transient events such as GRBs are even more challenging compared to persistent sources.
In this case, the difficulties of polarization measurements are compounded for two reasons: 
1). their duration is short ranging from few seconds to at most a few minutes, 
2). their occurrence is random in both space and time. All X-ray polarimeters measure the azimuthal distribution of scattered photons or photo-electrons with respect to a reference direction. 
Since GRBs occur at varying angles with respect to reference direction of large field of view GRB detector, the polarization measurements are highly prone to systematic effects. 
The short duration of GRBs results in a limited number of photons. Thus, measuring polarization of the GRB prompt emission is an ambitious task, 
and ideally requires a detector with large area to collect sufficient number of photons. 
Since the intrinsic polarization may be high it is often possible to achieve interesting results even with a small number of photons, if systematic effects are well understood. 

The GRB polarimetric studies provide a unique opportunity to address the nature of the magnetic fields close to the relativistic jet launching site \cite {toma08,lundman18}.
Based on this promising capability to study the central engines of GRBs, there have been many attempts to measure the polarization of prompt emission, 
both with dedicated polarimeters and instruments primarily designed for non-polarimetric observations. 
The first polarization measurement of GRB prompt emission was made in 2004 with 
the RHESSI instrument \cite {mcconnell02}, which is designed for solar hard X-ray spectroscopy. 
There has, however, been some controversy about these results \cite {rutledge04,wigger04}. 
Subsequently, there were reports of GRB polarization with INTEGRAL SPI \cite {mcgltnn07,mcglynn09} and IBIS \cite {gotz09,gotz13,gotz14}. 
There have also been attempts to measure the polarization of GRBs detected by BATSE onboard CGRO \cite {wills05}.
The GAP instrument onboard Japanese IKAROS mission was the first dedicated GRB polarimeter designed and calibrated for polarization measurements. 
GAP was launched in 2010 and detected polarization of three GRBs \cite{yonetoku11,yonetoku12}. 
COSI, a balloon-borne Compton spectrometer and imager, reported the polarization of a long GRB \cite {lowell17}. 
The second dedicated GRB polarimeter, POLAR \cite {orsi11}, was launched in 2016 onboard Chinese space-station Tiangong 2 and had a 7 months of 
operational lifetime, which measured polarization of 19 GRBs \cite {zhang19,kole20}. 
Summaries of GRB polarization measurements by various instruments can be found in McConnell (2017)~\cite{McConnell17}, Chattopadhyay (2021)~\cite{chattopadhyay21} and Gill et al. (2021)~\cite{gill21}.

The Cadmium Zinc Telluride Imager (CZTI )\cite{bhalerao17} onboard AstroSat is a hard X-ray 
imaging spectroscopy instrument, which is also capable of measuring 
X-ray polarization ~\cite{vadawale15,chattopadhyay14}. CZTI measured the polarization of 11 GRBs during its first year of operation~\cite {chattopadhyay19}.
	Recently, Chattopadhyay et al.(2022)\cite{chattopadhyay22} presented a catalog of polarization measurements of 20 GRBs selected with more stringent criteria on fluence and minimum detectable 
polarization from all the GRBs observed by CZTI during first five years of its operation.
GRB polarimetry of such a large sample with definite polarization measurements would help distinguish between various prompt emission models \cite {toma08}. 
Thus, polarization measurements with CZTI can significantly contribute to the study of GRB physics, if a high confidence in the measurements is established.

Most GRB polarization measurements, including those with CZTI, have a heavy reliance on Geant4 simulations.
Additionally, GRB polarimetry results have historically been plagued with large systematics and a number of controversies\cite{coburn03,rutledge04,wigger04,mcconnell17Book}.
In this context, experimental validation of the off-axis polarization sensitivity is vital.
In the case of CZTI, the pre-launch experimental validation involved only on-axis polarimetry~\cite{vadawale15}, and bridging this gap is the basic motivation behind the work presented here. 

In this paper we present experimental validation of the hard X-ray off-axis 
polarization measurement capability of the pixellated CZT detectors as well as 
the efficacy of the Geant4 simulations that are required for the polarization analysis. Here, we use both template fitting method and modulation curve fitting method for polarisation analysis.
Section \ref {sect:cztigrb} provides an overview of the GRB polarimetry with CZTI and the analysis methods. 
Section \ref{sect:exp} describes the experimental set-up used as well as the details of Geant4 simulations. 
Section \ref{sect:results} discusses the results of experiments and simulations. 
Finally, Section \ref{sect:Summary} summarizes the key findings of the present work. 


\section{GRB Polarimetry with AstroSat CZTI: Analysis methods}
\label{sect:cztigrb}
The CZTI instrument onboard AstroSat\cite{singh14} is designed primarily for hard X-ray imaging and spectroscopy up to 100 keV. As the coded mask and other support structures of CZTI become increasingly transparent at energies above 100 keV, CZTI can also act as a wide field of view GRB detector\cite{rao16} sensitive to polarization at these higher energies. Additionally, CZTI can be used for measurement of polarization for bright steady-state sources such as Crab\cite{vadawale18}.
Since the launch of AstroSat in September 2015, CZTI has detected more than 450 GRBs \cite{cztigrb}. Although GRBs span only a short duration compared to observations of bright persistent sources, GRBs are often detected well above the average background and hence the signal to background ratio is often much better for GRBs than for persistent sources. Thus, CZTI can also be used for the polarization measurements of bright GRBs~\cite{chattopadhyay19}.

Polarization measurements by CZTI are based on the principle of Compton scattering. When a high energy photon interacts in one of the CZT pixels by Compton scattering, it deposits a part of its energy in the scattering pixel. The scattered photon may escape the CZT detector, or get absorbed in the same pixel, or in the surrounding pixels. The first two cases result in single-pixel events, whereas in the third case, the scattered photon is absorbed in a nearby pixel, and the event is recorded as a double-pixel event. The angle between the scattering and absorbing pixel with respect to the instrument axis gives the approximate azimuthal scattering angle. Considering a 3 $\times$ 3 matrix of pixels with the center pixel being the scatterer, we can create a histogram with 8 bins using the absorbed events in the outer pixels. For polarimetric analysis with CZTI, we consider double-pixel events recorded within a coincidence time window of 40$~\mathrm{\mu s}$ in adjacent pixels. Compton events are further identified from these events based on the ratio of the energies deposited in the scattering to absorbing pixels~\cite{chattopadhyay14}.
\subsection{Modulation Curve Fitting Method}
From the scattering angles of the Compton events, an azimuthal histogram (referred as ASAD - \textit{Azimuthal Scattering Angle Distribution} hereafter) is generated.
This raw ASAD shows an inherent asymmetry even for unpolarized photons due to the difference in the solid angles subtended by the edge and the corner pixels.
This geometric effect is corrected by normalizing the observed raw ASAD with the ASAD of unpolarized photons obtained from Geant4 \cite {agostinelli03} simulations. The modulation present in this corrected ASAD contains the polarization information, which is inferred by fitting the modulation curve with a cosine (cos$^{2}\phi$ or cos2$\phi$) function. The amplitude of the fitted function determines the modulation amplitude ($\mu$). 
The polarization fraction (PF) is obtained by normalizing the best fit modulation amplitude by the modulation factor $\mu_{100}$, which is determined through simulations of the source with a 100\% polarized beam having the same energy spectrum, incident direction, and polarization angle as that of the observed ASAD. The polarization angle (PA) is obtained from the best fit of the phase of the modulation. This \textit{modulation curve fitting} approach is followed in the polarimetric analysis of GRBs presented in Chattopadhyay et al. (2019)\cite{chattopadhyay19} and Chattopadhyay et al. (2022)\cite{chattopadhyay22}. 

\begin{figure}[H]
\centering
\includegraphics[width=\linewidth]{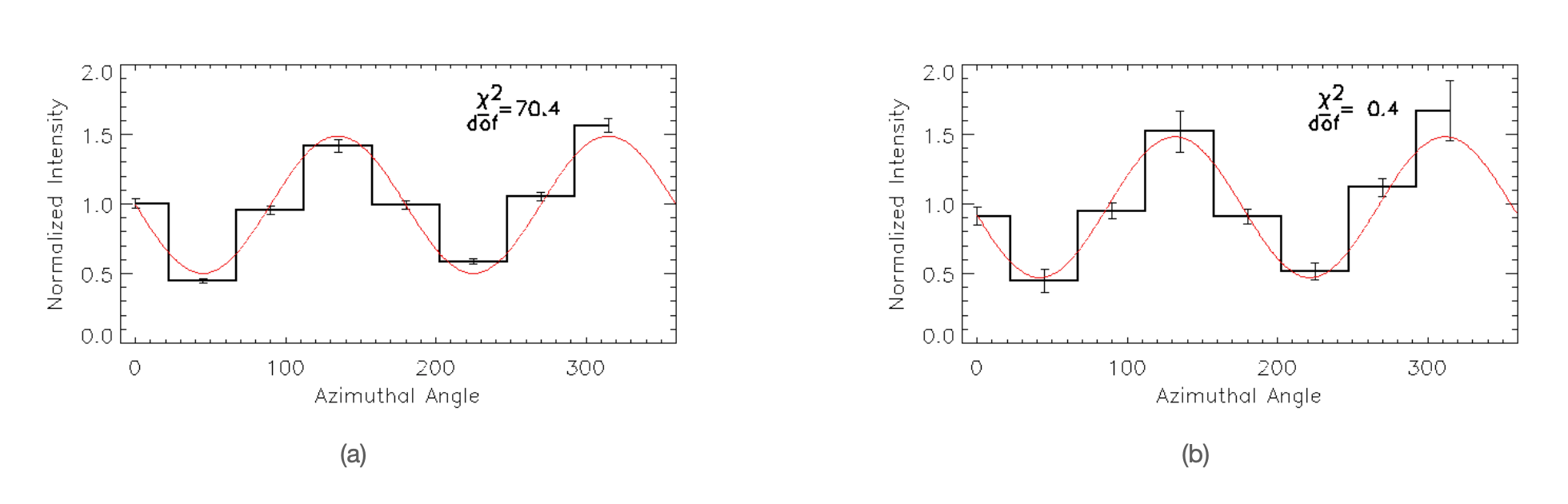}
\caption{\label{fig:modCurve} Simulated ASADs (normalised with unpolarised ASADs) with a best fitted cosine function for 100\% polarised photons corresponding to incident polar angle of 45$^\circ$ and PA 45$^\circ$. The reduced $\chi^2$ value clearly indicates the departure from the cosine nature. This departure is increasingly prominent at higher incident angles. (a) Shows the simulation ASAD for $\sim$ 0.4 million events, whereas (b) Shows the simulation ASAD for 10,000 events. Note: The error bars are obtained considering Poisson statistics on the events.}
\end{figure}

\begin{figure}[H]
\centering
\includegraphics[width=\linewidth]{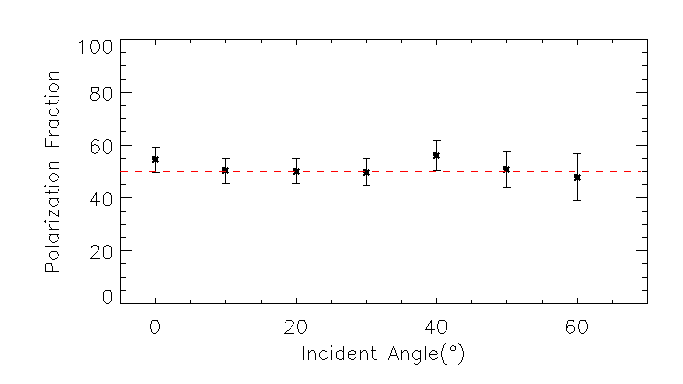}
\caption{\label{fig:PFdist} PF measurements by modulation curve fitting for 50\% polarised ASADs generated from polarised and unpolarised ASADs, for different incident angles. The red dotted line shows the 50 \% PF level.}
\end{figure}

While the modulation curve obtained by normalizing with
the unpolarized ASAD allows easier visualization of 
the presence of polarization signature, studies have
shown that inference of PF and PA by modulation curve
fitting is prone to systematic effects when used for
off-axis X-ray polarimetry\cite{muleri14}. Muleri (2014)
has shown that the geometry corrected ASAD does not
follow a cosine distribution when the photons are
incident at off-axis angles as the ASAD samples
different ranges of polar scattering
angles~\cite{muleri14}. We examined this effect in the case of CZTI by carrying out Geant4 simulations of
polarized and unpolarized parallel beams incident at different polar angles. (These simulations are similar to the simulations done for generating template library (Section \ref{sect:simgrid}).) ASADs obtained from the polarized simulations were normalized using the
unpolarized ASADs and were fitted with a cosine function. It was observed that the ASADs generated
for a large number of Compton events showed systematic departure from a cosine function, which is increasingly prominent at large off-axis angles. An example ASAD at incident angle of 45$^{\circ}$ is shown in Figure \ref{fig:modCurve}a.
However, the number of Compton
events in the case of actual GRBs is rather limited, often much less than 10000. Thus, we repeated the same exercise, but with only 10000 Compton events randomly
selected from the simulations. In this case, as the uncertainties are larger, the ASADs are statistically consistent with a cosine function as shown in Figure~\ref{fig:modCurve}b.
Additionally, we also verified that the PF values from modulation curve fitting provide accurate measurements of actual PF of ASADs generated by partially polarized beams.
This verification was carried out by creating 50\% polarised ASADs ($\sim$10,000 events) from the polarised and unpolarised ASADs. These partially polarised ASADs are also fitted with modulation curve and obtained PF. The measurements of PF for different incident angles are given in Figure~\ref{fig:PFdist}. It can be seen from this figure that the true PF value is recovered for all the incident angles.  

Although this analysis confirms that the assumption of cosine nature of the corrected ASAD does not have any major impact on CZTI GRB polarimetry with limited number of photons, we present an alternate analysis method, which does not involve such assumptions, to infer PF and PA.     
\subsection{Template Fitting Method}
In this alternate method we directly compare the observed raw ASAD (without normalizing by unpolarized ASAD) with the expected ASAD obtained from Geant4 simulations for a grid of PFs and PAs. A similar approach is followed in the analysis of GRB polarization from the POLAR experiment \cite {zhang19}. 
This analysis requires a library of ASADs for all possible
PFs and PAs from a given incident
direction, generated using Geant4 mass model simulations, 
as discussed further in section \ref {sect:simgrid}. 
The observed ASAD is directly compared with each of the simulated ASADs of the library to obtain the 
$\chi^{2}$ value between the measured and simulated 
modulation as given by:

\begin{equation}
\label {eqn:4}
{\chi^{2}    = \sum_{i=1}^{8} \frac{(M_{obs,i} - M_{sim,i})^{2}}{(\sigma_{obs,i}^{2} + \sigma_{sim,i}^{2})}}\ \  
\end{equation}
and $\Delta\chi^{2}$ is defined as,
\begin{equation}
\label {eqn:5}
{\Delta \chi^{2}    = \chi^{2} - \chi^{2}_{min}}\ \  
\end{equation}
\noindent Here, M$_{obs,i}$ and M$_{sim,i}$ 
are the observed and simulated number of counts in each of the 8 bins of the ASAD, and $\sigma$$_{obs,i}$ and 
$\sigma$$_{sim,i}$ are the corresponding errors, and $\chi^{2}_{min}$ is the minimum value of $\chi^2$ over the full range of PF and PA. The PF and PA corresponding to $\chi^{2}_{min}$ are reported as the best fit values. The 
error estimates for PF and PA are obtained from the
two parameter confidence levels for the $\chi^{2}$ distribution
with $\Delta\chi^{2}$ = 2.28, 4.61 and 9.21 for the 
1$\sigma$, 2$\sigma$ and 3$\sigma$ confidence levels respectively.


In the present work, we primarily use the template fitting method for analysis of the experimental and simulation data as it does not involve any assumptions on the nature of the modulation. We also analyze the same data using the modulation curve fitting approach and 
compare the results from both methods. 
This is particularly important when the measurements are limited by the counting statistics, which is expected to be the case for astrophysical GRBs.

\section{Off-Axis Polarization Experiment using CZT Detector} 
\label{sect:exp}

In an earlier work, Vadawale et al.(2015)~\cite{vadawale15} have shown the capability of pixellated CZT detectors, used in AstroSat CZTI, in measuring the polarization of on-axis sources over an energy range of 100 -- 400 keV. In this work, we extend this to off-axis sources by carrying out experiments with a partially polarised X-ray source shining on the detector at different incidence angles. In this section, the details of the experiment and associated Geant4 simulations are discussed.

\subsection{Off-Axis polarization experimental set-up}

To carry out the present experiment, we developed a new detector readout system with a single CZT detector module following the same readout methodology employed in AstroSat CZTI.
The readout electronics records time-tagged list of events along with their interaction pixel and pulse height similar to the event list from CZTI. 
The CZT detector module used in the experiment is a spare from the same batch of flight module detectors used in CZTI and has similar performance in terms of efficiency, gain, and energy resolution. This CZT detector is a pixellated detector having 256 pixels of size 2.46 mm $\times$ 2.46 mm.
Firstly, we obtained long duration background measurements and identified seven pixels having higher noise. These pixels were disabled for all subsequent experiments. 
We then carried out experiments using radioactive sources $^{241}$Am, $^{133}$Ba and $^{109}$Cd to obtain the gain and energy resolution of individual pixels of the detector. 
The pixel-wise gain is used to obtain nominal energy of events from the measured pulse heights.

\begin{figure}[H]
\centering
\includegraphics[width=\linewidth]{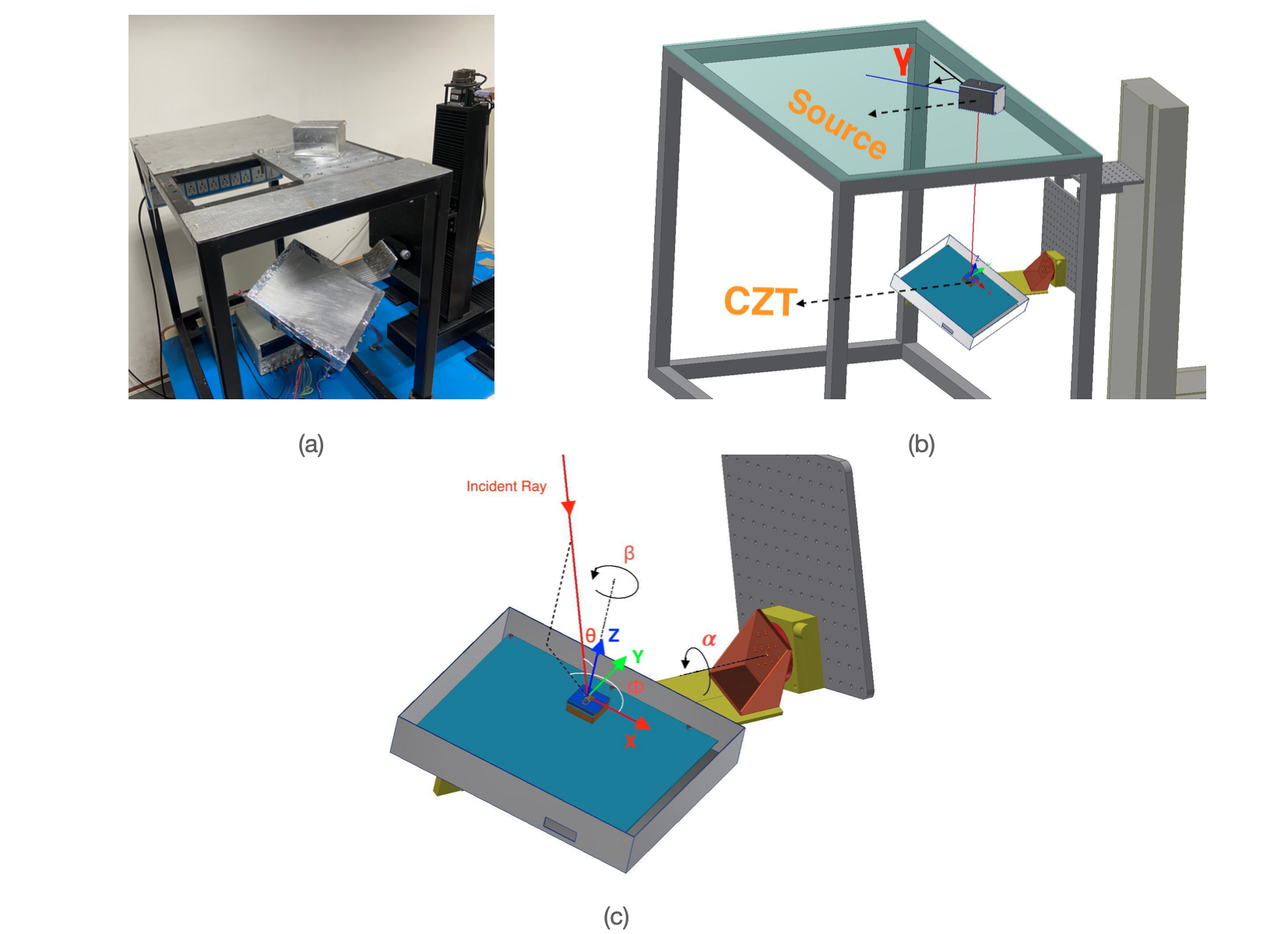}
  \caption
 {\label{fig:ExptCAD} The images show the experimental set-up 
for off-axis measurements and analogous representation of the same in CAD. 
(a) the full experimental set-up with partially polarised X-ray source mounted on a surface above the detector.
(b) the CAD model representation of full experimental set-up.
(c) the detector mount set-up and corresponding angular and coordinate system definitions.
}
\end{figure}

In order to illuminate the detector with partially polarised X-rays at different incidence angles, we developed an experimental setup as shown in Figure~\ref{fig:ExptCAD}.
The CZT detector along with the readout electronics board is enclosed in an aluminum box ( 28 x 33 x 5 cm) and is mounted on a 3-axis translation stage. 
The detector box mount has two angular degrees of freedom: angle $\alpha$ that corresponds to the polar angle can be changed by the rotational stage and angle $\beta$ that corresponds to the azimuthal angle can be varied by mounting the detector box at specific angles (0$^{\circ}$ and 30$^{\circ}$) on the bracket, as shown in Figure \ref{fig:ExptCAD}c.
The partially polarized X-ray source is mounted on top of a table above the detector. 
The translation stages are used to align the detector with the incident X-rays from the source. 
By changing the angles $\alpha$ and $\beta$, we can carry out the experiment at different polar polar ($\theta$) and azimuthal ($\phi$) angles of the incident X-rays with respect to the detector coordinate system (see Figure~\ref{fig:ExptCAD}c). 
The partially polarised X-ray source can also be rotated at different angles ($\gamma$) as shown in figure \ref{fig:ExptCAD}a to achieve different polarisation angles (PA). 
The PA mentioned here is defined in the sky plane normal to the incident direction along $\theta$ and $\phi$ and it is measured from local north towards east.   

\begin{figure}[H]
\centering
\includegraphics[width=\linewidth]{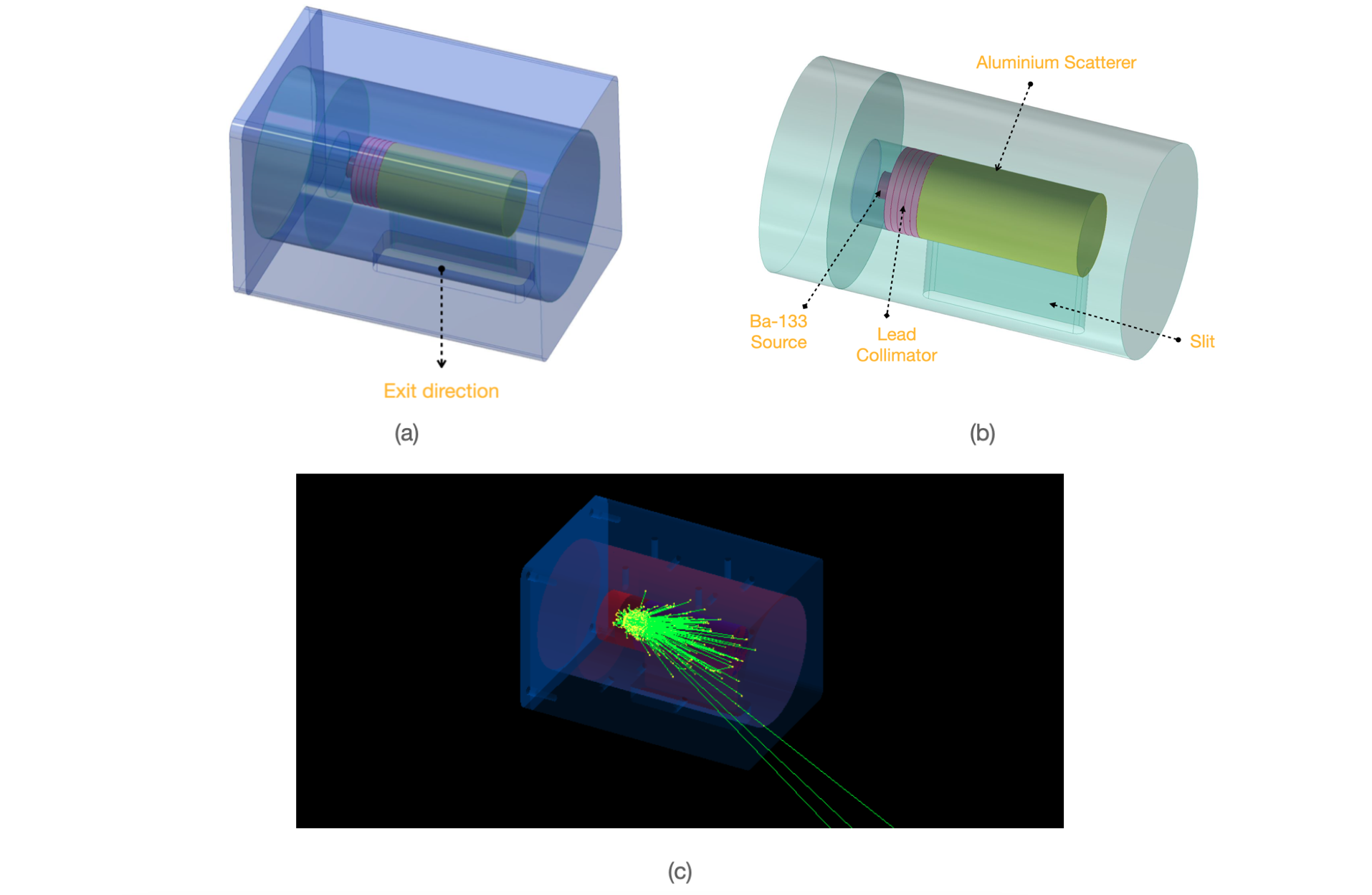}
\caption 
{\label{fig:sourceCAD} 
(a) The CAD model of the partially polarised X-ray source.
(b) The source assembly without outer aluminium container.
(c) The Geant4 simulated partially polarised X-ray source set-up. The incident $^{133}$Ba X-rays (green) can be seen getting scattered by the Aluminium cylinder inside the assembly.}

\end{figure}

We generate partially polarized X-rays by Compton scattering unpolarized X-rays from a disk-shaped $^{133}$Ba radioactive source (Figure~\ref{fig:sourceCAD}). 
A 6 cm long aluminum cylinder inside a 4 cm thick lead enclosure is used as a scatterer. A collimator of dimension 2 mm radius and 10 mm length is placed in front of $^{133}$Ba, allowing only a narrow beam to incident on the aluminium scatterer.
The scattered (partially polarized) X-rays exit the enclosure in one direction through a slit of 5 cm length and 2 mm width (Figure~\ref{fig:sourceCAD}b). It is to be noted that the emerging partially polarized beam is non-parallel in nature. 

As the incident photons from astrophysical sources are parallel, ideally, the experiments are to be conducted by shining a fully or even partially polarized parallel beam of hard X-rays with known PF and PA at the energies of interest, incident at different angles with respect to the detector plane. 
But it is practically impossible to obtain such parallel beam of hard X-rays for laboratory experiments. 
In order to approximate the emerging partially polarised beam to a parallel beam, the X-ray source is placed at maximum possible height (30 cm) such that signal to background ratio is high enough to carry out the experiments.

We carried out experiments for different polar ($\theta$), azimuthal ($\phi$), and polarization angles (PA). Data were acquired by placing incident beam at $\theta$ of 30$^{\circ}$, 45$^{\circ}$, 50$^{\circ}$, and 60$^{\circ}$ for some combinations of $\phi$ and PA. Long duration background exposures were also taken for each case. 
We also acquired data with an unpolarized beam of X-rays at different incidence angles. The $^{133}$Ba source is used without the scattering element to provide unpolarized radiation with its most prominent line at 356 keV.

For each measurement, data were recorded in the form of event lists and gain correction was applied to event pulse heights to obtain nominal energy of the events. 
From the pixel light curves, noisy pixels and flickering pixels were identified. We removed the events from these pixels, similar to the procedure followed in the data analysis pipeline of CZTI. 
From the clean event list, Compton double pixel events were selected and ASADs are generated as described in previous section. 
Similar analysis is carried out for background observations and background ASADs are subtracted to obtain final ASADs which are used in subsequent polarization analysis by template or modulation curve fitting methods.

\subsection{Geant4 simulations} 
\label{sect:sim} 
 
Geant4 simulations are an integral part of X-ray polarization analysis. 
In the case of polarization analysis with CZTI for X-rays incident on-axis, the details of Geant4 simulations were reported by Chattopadhyay et al.(2014)\cite{chattopadhyay14} and Vadawale et al.(2015)\cite{vadawale15}.
In the present work, we use a similar methodology for the Geant4 simulations including detector geometry, physics processes, event processing, and further data analysis to finally record the ASAD in the detector plane.
A pixellated CZT detector matching the dimensions of the actual detector is modelled in the simulations and the details of Compton double pixel events are recorded from the simulations for generating raw ASADs. 
We carry out two sets of Geant4 simulations with different source configurations with the CZT detector. 
In the first set, simulation of the complete experimental setup including the generation of the partially polarized source through scattering within the source assembly is carried out, placing the source at different orientations with respect to the detector exactly as done in the experiment. 
The second set of simulations are for the analysis of the data from both experiment and simulation of the experimental setup. 
Here, simulations with 100\% polarized and unpolarized X-rays incident on the detector at the same incident angle as that of experiment are carried out which are required for the analysis by both template fitting and modulation curve methods. Details of these two simulations are presented here.

\subsubsection{Simulation of the full experimental set-up}
\label{sect:simsetup} 

In the experiment, partially polarized X-rays are produced by scattering the X-rays from a $^{133}$Ba radioactive source using the assembly shown in Figure~\ref{fig:sourceCAD}. 
We model this assembly in Geant4 so that we can estimate the polarization fraction and angle of the scattered X-rays reaching the detector and compare the results of the simulation with the experiment. 
Various components of the partially polarized X-ray source assembly, namely, hollow lead cylinder with cap, collimator, source holder, aluminium holder, and aluminium scatterer, are defined in the exact dimensions by importing their CAD designs using the CADMesh interface for Geant4~\cite{poole12}. 
This direct import of the CAD design preserves the relative placement and orientation of various components of the source assembly. 
The complete assembly is then placed at different orientations with respect to the detector by defining appropriate rotation matrices. 
The X-rays from $^{133}$Ba source are defined using General Particle Source (GPS) as a disc placed at appropriate orientation emitting isotropically in the direction towards the aluminium scatterer. Energy spectrum of GPS includes all strong lines emitted by $^{133}$Ba with corresponding relative intensities.

For each configuration of $\theta$, $\phi$, and PA for which experiments are done, simulations are also carried out. 
As only a small fraction of source photons gets scattered by the aluminium scatterer and reach the CZT detector, the number of incident photons for each simulation is chosen to be 10$^{10}$ such that the number of Compton events detected in CZT detector is $\sim$ 10,000. 
In addition to the raw ASAD of the Compton events, we also save the momentum vector and polarization vector of the photons incident on the detector. 
The energy spectrum and the distribution of incident directions are computed from the momentum vectors of the incident photons. 
The incident energy spectrum obtained this way is used for the simulations described in the next section. 
From the polarization vectors, we calculate Stokes parameters for individual photons. 
The summation of these Stokes parameters gives the estimation of the actual polarization fraction and sky plane polarization angle of the incident beam on the detector. 
These true values of PF and PA are used for the comparison with the measured values from raw ASADs obtained from experiment as well as the simulations of the experiment setup.

\subsubsection{Generation of the ASAD library}
\label{sect:simgrid}

For polarimetric analysis with either template or modulation curve fitting method, we require simulations with polarized and unpolarized X-rays with the same incident spectrum and direction as discussed in Section~\ref{sect:cztigrb}. 
For each experimental configuration, simulations are carried out where the CZT detector is directly illuminated with a parallel beam of 100\% polarized and unpolarized X-rays incident at the respective polar($\theta$) and azimuthal($\phi$) angles. 
Energy spectrum of the primary photons is set to the incident spectrum obtained from the simulation of the experimental setup as discussed in previous section. 
For each incident direction, the simulations are performed for 100\% polarized beam for all PAs from 0$^\circ$ -- 180$^\circ$ at steps of  5$^\circ$. 
Simulations are carried out for 10$^{8}$ incident photons such that $\sim$300,000 -- 600,000 Compton events are detected, which are used to generate the raw ASADs.

To obtain the library of ASADs over a grid of PF and PA as required for template fitting, randomly selected Compton events from polarized simulations are fractionally added to the unpolarized 
Compton events for every 1\% polarization step. ASADs obtained at 5$^\circ$ PA are interpolated to obtain ASADs at every degree of PA, resulting in a library over a grid of 180 PA bins  $\times$ 101 PF bins, which is used to 
calculate $\chi^2$ at each point in the PF - PA  grid for the observed raw ASADs. 

For normalizing the ASADs in modulation curve fitting, unpolarized ASADs obtained from these simulations are used. 
Modulation amplitude ($\mu_{100}$) as well as phase angle ($\phi_{o,100}$) for 100\% polarized photons is obtained for each sky plane PA from the ASADs of polarized simulations. 
The $\mu_{100}$ value corresponding to the phase angle of the observed modulation is used to calculate the polarization fraction.
Sky plane polarization angle corresponding to the phase angle of observed modulation is reported as PA.

\section{Results and Discussion } 
\label{sect:results} 

For each of the experimental configurations ($\theta$, $\phi$, and PA), we analyse the results of simulations of the experimental setup. 
True polarization fraction and polarization angle of the beam incident on the detector estimated from the simulations as discussed in previous section are listed for each of the configurations in Table~\ref {tab:1}. 
In all cases, the incident beam has similar polarization fraction of about 62\%. 
In our analysis of ASADs, statistical errors on the each bin of ASADs are obtained considering Poisson statistics on the events and propagated appropriately.
As the incident beam from the partially polarized source is not plane parallel, a systematic error of 3\%, estimated from the departure of the beam from parallel nature, is added to the statistical uncertainties. 

In the analysis of experimental ASADs, we had to account for charge sharing effect. Chattopadhyay et al. (2022)\cite{chattopadhyay22}, for a sample of twenty bright GRBs, showed that the ratio of Compton to Single Pixel events is consistently higher in experimental data than in simulation results (see Figure 1 of Chattopadhyay et al., 2022\cite{chattopadhyay22}). 
This effect is attributed to charge sharing between the adjacent pixels which can mimic a Compton event and consequently add to the number of true Compton events. 
After correcting for the number of Compton events and Single Pixel events in simulation using an empirical model for charge sharing, the ratio of Compton to single pixel events was found to agree closely with the observed ratios for all the twenty GRBs. 
We have noticed the same effect in the experimental ASADs when compared to the simulated ASADs.
Since the simulated ASADs are used for template fitting or geometry correction (in the case of modulation curve fitting), it is important to account for this effect in the simulated distribution which is done by scaling the ratio of edge and corner bins of unpolarized and polarized simulations with the mean of the edge and corner bins of the experimental raw ASAD, as suggested in Chattopadhyay et al. (2022)\cite{chattopadhyay22}. These corrected simulation ASADs are used in further polarimetric analysis of experimental ASADs. 

The raw ASADs, obtained from the simulations, are analysed using template fitting method.
Estimated PF$_{sim}$ \& PA$_{sim}$ corresponding to the $\chi$$^{2}$$_{min}$ along with corresponding 1$\sigma$ error bars for all cases are listed in Table~\ref {tab:1}. 
It can be seen that the measured PF and PA values match closely with the incident true PF and PA values, with only a couple of configurations deviating by more than 1 sigma, showing the efficacy of the off-axis Compton polarimetry with pixellated detectors.

\begin{figure}[H]
\centering
\includegraphics[width=\linewidth]{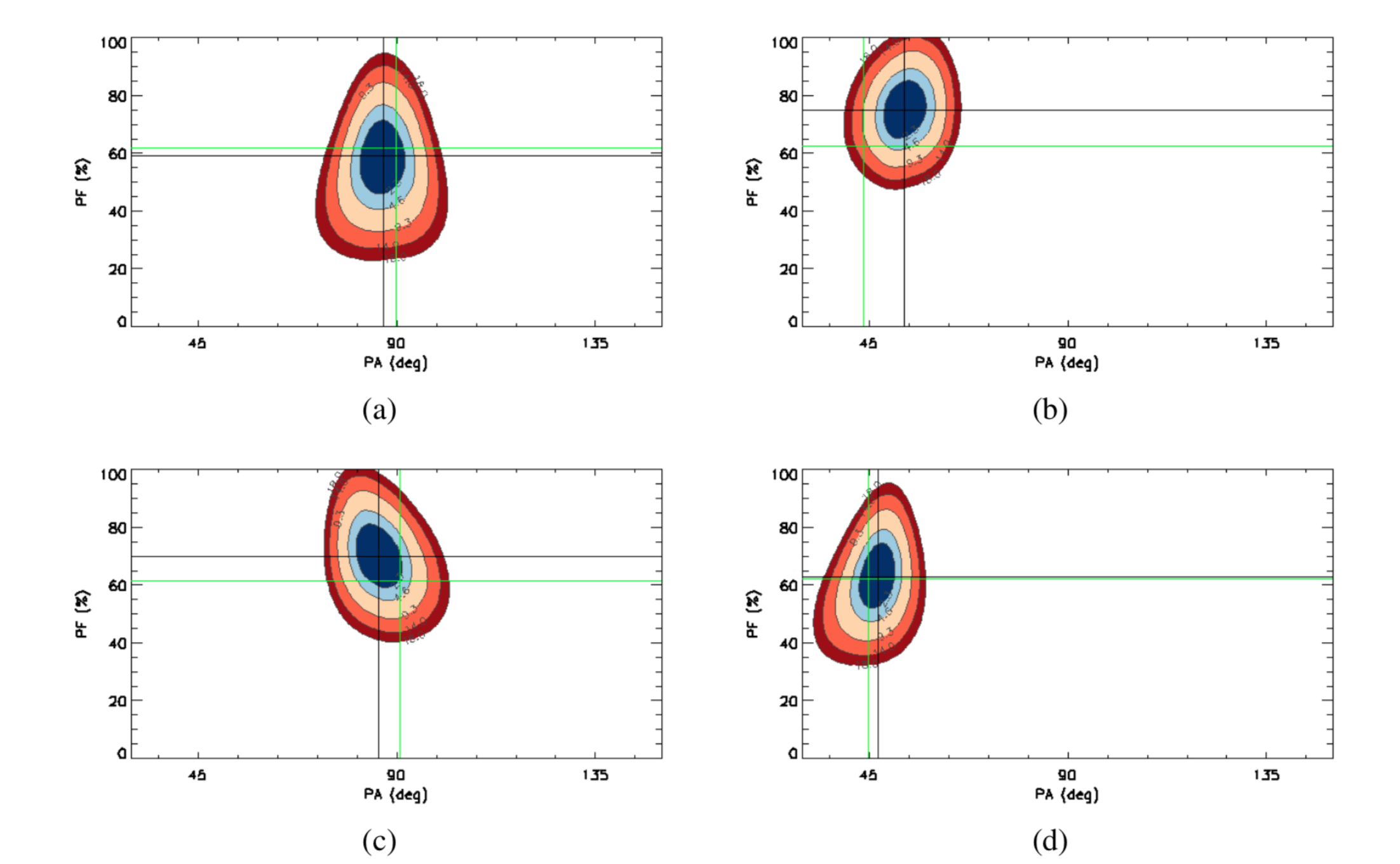}
\caption{The plots show the confidence contours
from experiment data obtained for:
(a) $\theta$ = 30$^\circ$, $\phi$ = 0$^\circ$, and PA of 90$^\circ$
(b) $\theta$ = 30$^\circ$, $\phi$ = 0$^\circ$, and PA of 45$^\circ$. 
(c) $\theta$ = 30$^\circ$, $\phi$ = -30$^\circ$, and PA of 90$^\circ$. 
(d) $\theta$ = 30$^\circ$, $\phi$ = -30$^\circ$, and PA of 45$^\circ$. 
The black lines shows the PF, PA values corresponding to the minimum $\chi^2$ value, whereas green lines represent the incident PF, PA values. The colour contours represents 1$\sigma$, 2$\sigma$, 3$\sigma$, 4$\sigma$ and 5$\sigma$ confidence levels.}
\label{fig:30degContours}
\end{figure}

\begin{figure}[H]
\centering
\includegraphics[width=\linewidth]{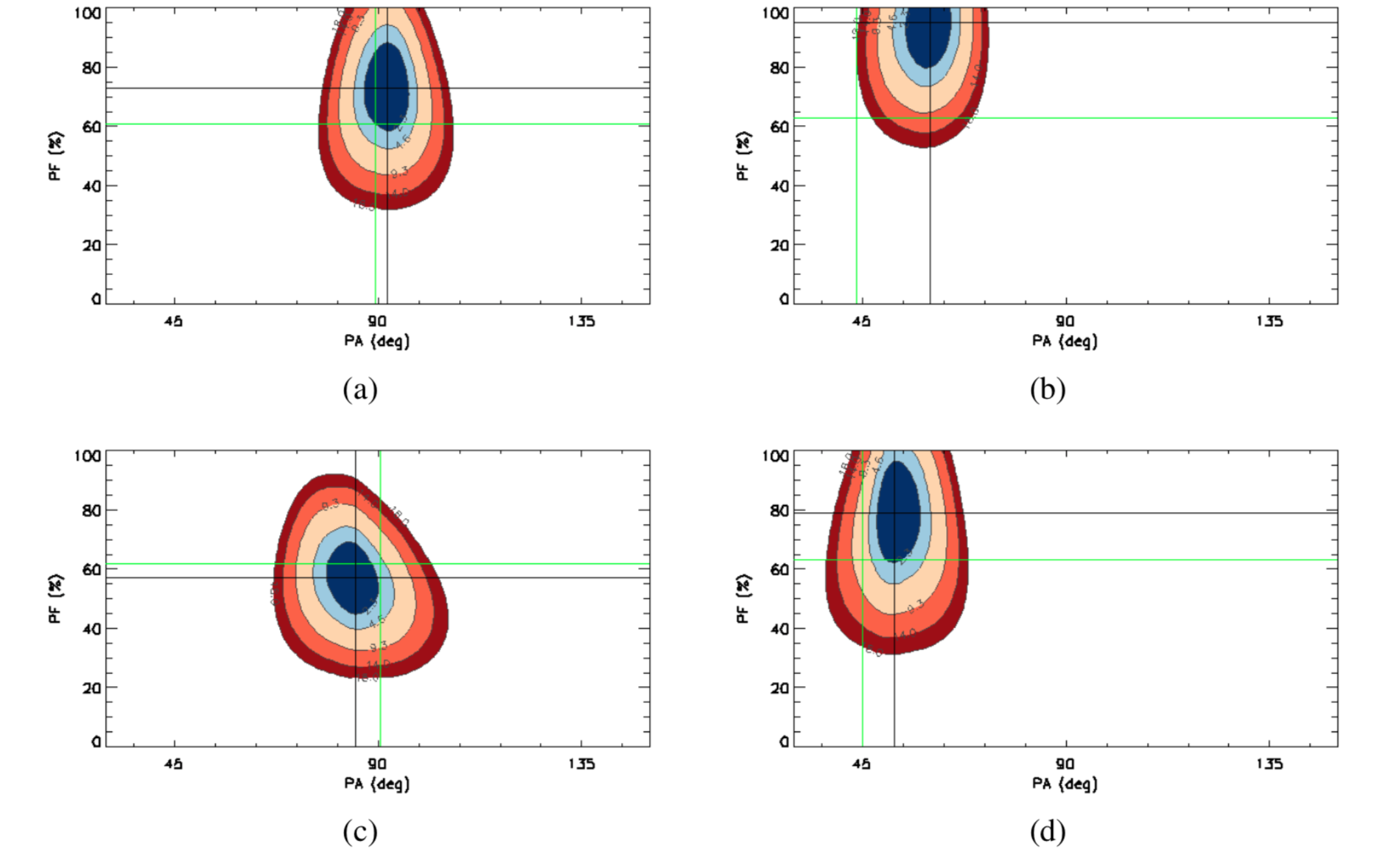}
\caption{The plots show the confidence contours 
from experiment data obtained for: 
(a) $\theta$ = 45$^\circ$, $\phi$ = 0$^\circ$, and 
PA of 90$^\circ$ 
(b) $\theta$ = 45$^\circ$, $\phi$ = 0$^\circ$, and PA of 45$^\circ$. 
(c) $\theta$ = 45$^\circ$, $\phi$ = -30$^\circ$, and PA of 90$^\circ$. 
(d) $\theta$ = 45$^\circ$, $\phi$ = -30$^\circ$, and PA of 45$^\circ$. 
The black lines shows the PF, PA values corresponding to the minimum $\chi^2$ value, whereas green lines represent the incident PF, PA values. The colour contours represents 1$\sigma$, 2$\sigma$, 3$\sigma$, 4$\sigma$ and 5$\sigma$ confidence levels.}
\label{fig:45degContours}
\end{figure}

\begin{figure}[H]
\centering
\includegraphics[width=\linewidth]{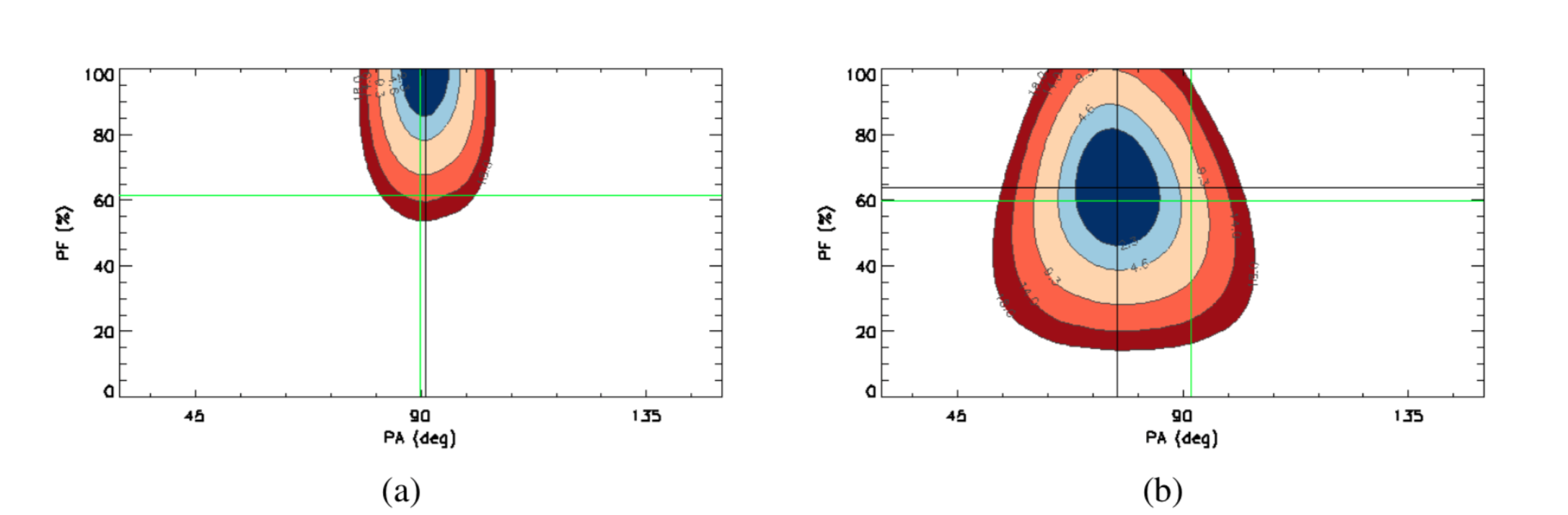}
\caption{The plots show the confidence contours
from experiment data obtained for:
(a) $\theta$ = 50$^\circ$, $\phi$ = 0$^\circ$, and
PA of 90$^\circ$.
(b) $\theta$ = 50$^\circ$, $\phi$ = -30$^\circ$, and
PA of 90$^\circ$.
The black lines shows the PF, PA values corresponding to the minimum $\chi^2$ value, whereas green lines represent the incident PF, PA values. The colour contours represents 1$\sigma$, 2$\sigma$, 3$\sigma$, 4$\sigma$ and 5$\sigma$ confidence levels.}
\label{fig:50degContours}
\end{figure}

\begin{figure}[H]
\centering
\includegraphics[width=\linewidth]{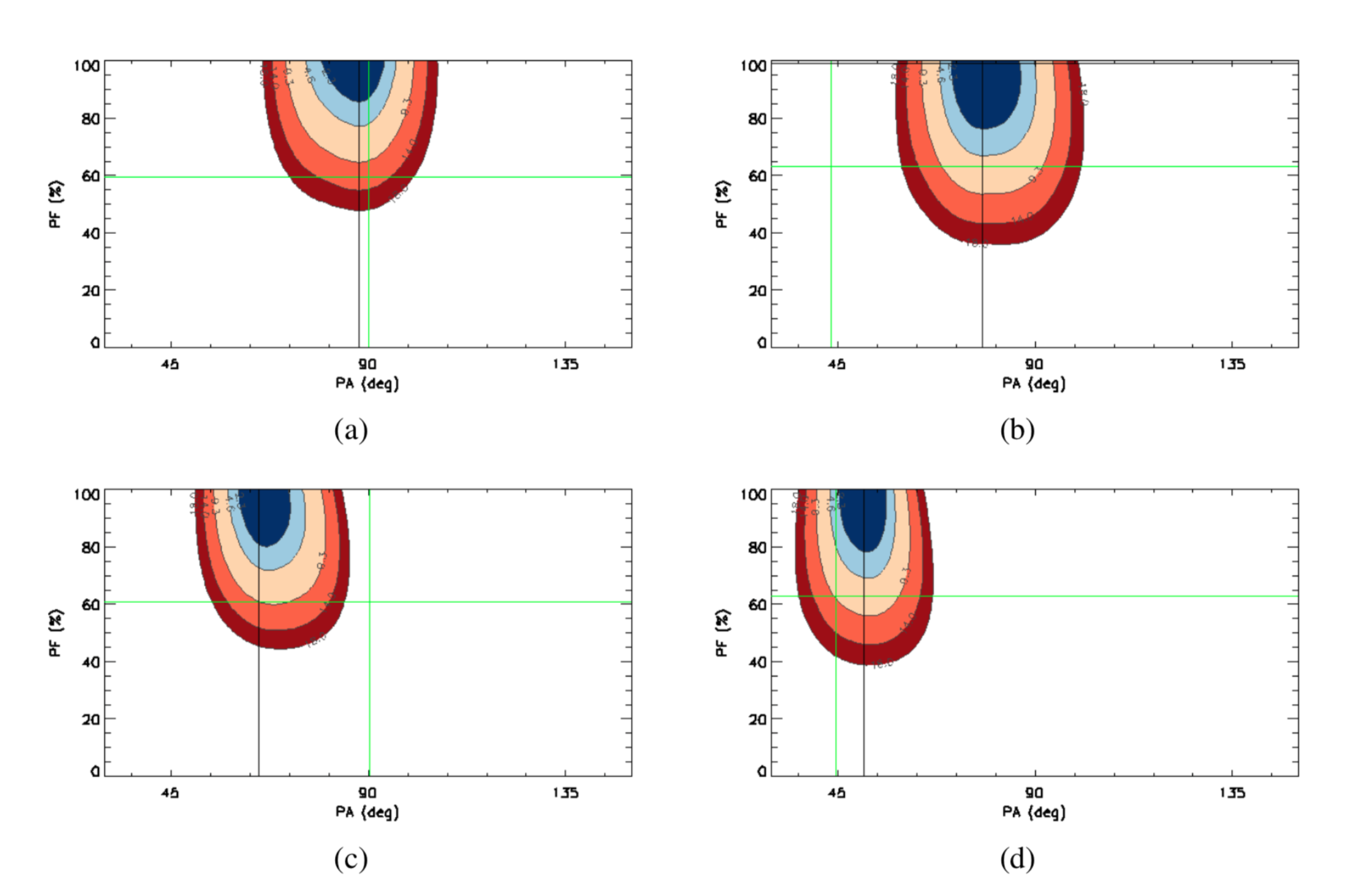}
\caption{The plots show the confidence contours
from experiment data obtained for:
(a) $\theta$ = 60$^\circ$, $\phi$ = 0$^\circ$, and
PA of 90$^\circ$.
(b) $\theta$ = 60$^\circ$, $\phi$ = 0$^\circ$, and
PA of 45$^\circ$.
(c) $\theta$ = 60$^\circ$, $\phi$ = -30$^\circ$, and PA of 90$^\circ$. 
(d) $\theta$ = 60$^\circ$, $\phi$ = -30$^\circ$, and PA of 45$^\circ$. 
The black lines shows the PF, PA values corresponding to the minimum $\chi^2$ value, whereas green lines represent the incident PF, PA values. The colour contours represents 1$\sigma$, 2$\sigma$, 3$\sigma$, 4$\sigma$ and 5$\sigma$ confidence levels.}
\label{fig:60degContours}
\end{figure}

The raw ASADs, obtained from the experiments, are then analysed using template fitting method.
For each experimental configuration, the ASADs are fitted with template library taking into account statistical and systematic uncertainties similar to the analysis of simulation data to obtain PF and PA corresponding to $\chi$$^{2}$$_{min}$ and $\Delta\chi^{2}$ array over the grid of PF and PA. 
Confidence contours for PF and PA obtained from this analysis for different experimental configurations are shown in Figures \ref{fig:30degContours} to \ref{fig:60degContours}. 
The PF and PA values obtained for each configuration along with corresponding 1$\sigma$ error bars are listed in Table~\ref{tab:1}.
We find that the experimental measurements are also consistent with the incident PF and PA values, except for some cases at large off-axis angles of 50-60$^{\circ}$, which we discuss later.

We also analysed the simulation and experimental data using the modulation curve fitting method. ASADs normalized with unpolarized simulated ASADs are fitted with cosine function to obtain the modulation amplitude ($\mu$) and phase angle $\phi_{o}$. 
Amplitude for 100\% polarization ($\mu_{100}$) and sky plane PA corresponding to the best fit  $\phi_{o}$ are computed by interpolating the values available over the grid of PAs. 
PF and PA measured in this way are also shown in Table~\ref{tab:1} for all cases of experiment and simulations. 
It can be seen that the measurements by modulation curve fitting are consistent with the measurements using template fitting. 
These results again show that the systematic effects in modulation curve fitting used for off-axis polarimetry is negligible when the statistical uncertainties are large as in the case of GRB polarimetry.

\begin{figure}
\centering
\includegraphics[width=\linewidth]{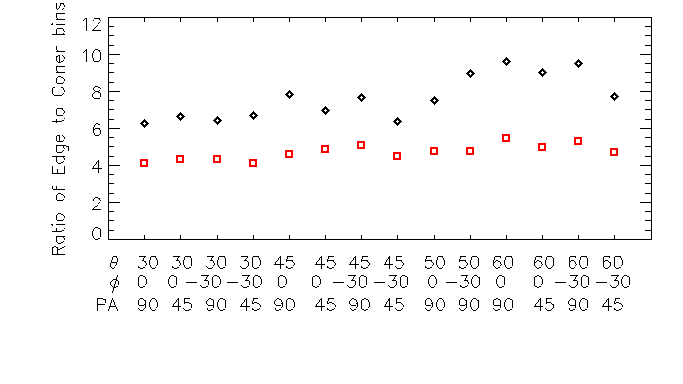}
\caption
{\label{fig:EdgeCorner} The ratio of total events in edge bins to corner bins of simulation and experimental ASADs for different experimental configurations are shown. The black points corresponds to experimental data whereas red points corresponds to simulations.} 
\end{figure}

From Figure~\ref{fig:60degContours} and Table~\ref{tab:1}, it can be seen that the PF measured from experiment for incidence angles close to 60$^{\circ}$ tend to be higher than the PF of the incident beam. 
Since a similar trend is also seen in the measured PF from simulations of the experimental setup, we believe that the approximation of the actual diverging beam from the polarized source to a parallel beam does not hold at higher off-axis angles. 
We also notice that the differences between the ratios of edge to corner bin events in experiments and simulations are slightly higher than the average at angles above 45$^\circ$ as shown in Figure \ref{fig:EdgeCorner}. 
There might be some inherent angle dependence of charge sharing effect that could lead to this systematic effect. Understanding these effects requires detailed modelling and more experimental validations at far off-axis angles which we plan to carry out in the future.   
It is important to note that both these effects (divergence and systematic above 45$^\circ$) do not have any significant effect in case of CZTI GRB polarimetry analysis. 
Particularly, the divergence of the beam is not an issue in the case of GRB analysis where the incident rays are parallel. 
Even the systematic effects at large off-axis angles are expected to be minimal compared to the statistical uncertainties in case of CZTI GRBs, which, in most cases, register around 1000-3000 Compton events.

\begin{figure}
\centering
\includegraphics[width=\linewidth]{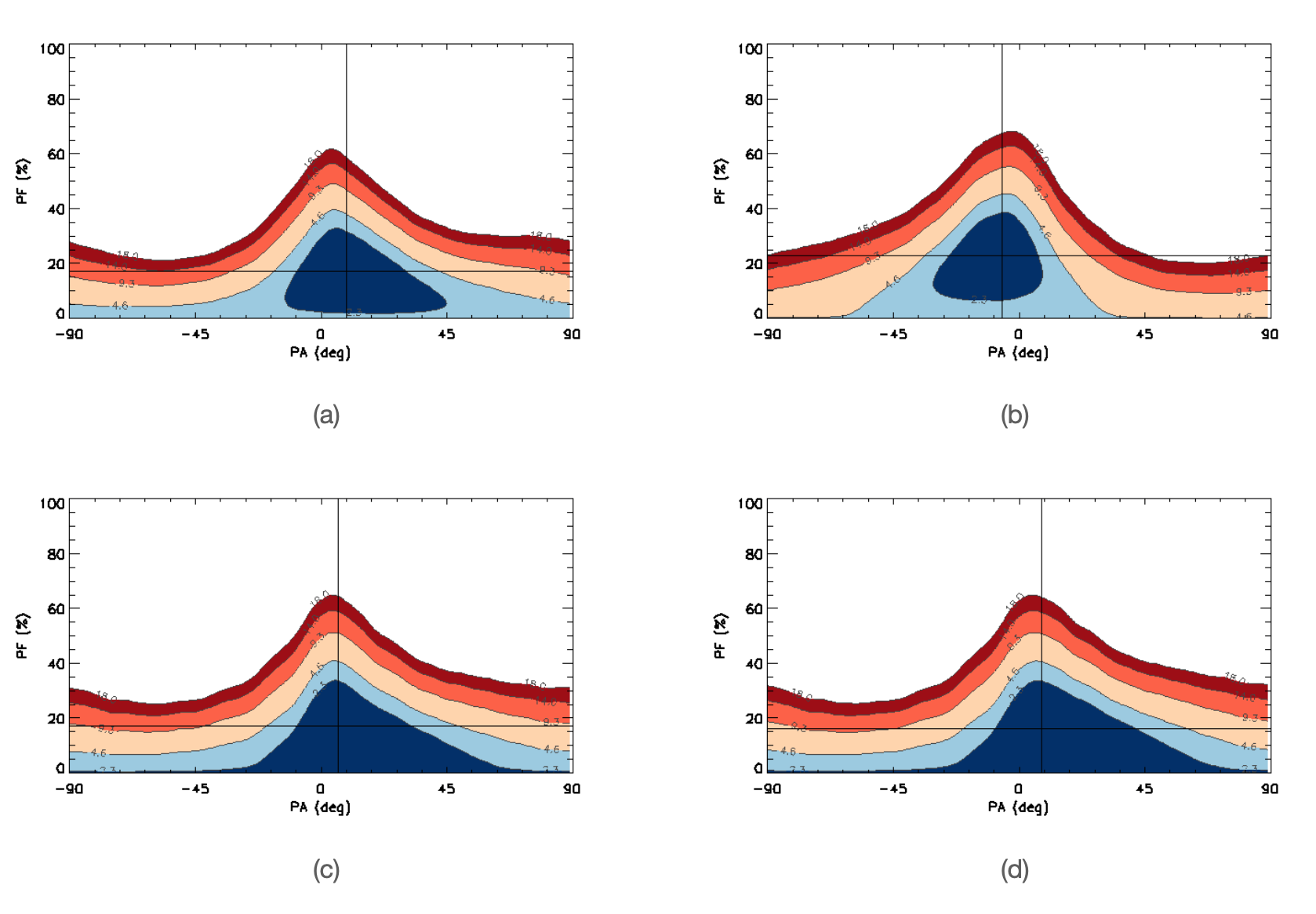}
\caption{The plots show the confidence contours
from experiment data obtained for unpolarized incident radiation:
(a) $\theta$ = 30$^\circ$, $\phi$ = 0$^\circ$, 
(b) $\theta$ = 45$^\circ$, $\phi$ = 0$^\circ$, 
(c) $\theta$ = 50$^\circ$, $\phi$ = 0$^\circ$, 
(d) $\theta$ = 60$^\circ$, $\phi$ = 0$^\circ$
The black lines shows the PF, PA values corresponding to the minimum $\chi^2$ value. The colour contours represents 1$\sigma$, 2$\sigma$, 3$\sigma$, 4$\sigma$ and 5$\sigma$ confidence levels.}
\label{fig:UnPolcontours}
\end{figure}

We also obtained measurements with unpolarized radiation incident on the detector at different off-axis angles 
for the same number of Compton events as in the polarized experiments ($\sim$10000).
The obtained ASADs are fitted with template library generated with photons of the same incident spectrum and direction. 
Results of this analysis are shown in Figure~\ref{fig:UnPolcontours}. It can be seen that the unpolarized radiation starts mimicking polarized X-rays at higher angles,
which means that the sensitivity is reduced with increasing polar angle of incidence. This is due to
the fact that at different incidence angles, the ASADs recorded in the detector plane
sample different polar and azimuthal scattering angles as governed by the scattering cross-section.

To get a limit on the polarimetric sensitivity at different incidence angles, 
we estimate the Minimum Detectable Polarization (MDP) using polarised and unpolarised simulations at different angles.
The standard method for MDP calculation assumes an unmodulated azimuthal distribution for unpolarized radiation. However, in the case of off-axis measurements, this assumption is not completely valid and thus it is not appropriate for the MDP calculation of off-axis polarimeters. 
So, instead, we follow an alternate approach of obtaining MDP from its basic definition using simulations. MDP is defined as the minimum polarization fraction required so that the probability of observed modulation being attained by unpolarized photons by chance is less than 1\%. In other words, polarization measurements of an unpolarized source will result in PF less than MDP 99\% of the time. We generate several realizations of unpolarized ASADs and then fit them to obtain measurements of PF and PA. MDP is then estimated as the 99th percentile of the distribution of measured PFs marginalized over different ranges of PA.

\begin{figure}
\begin{center}
\includegraphics[width=1\linewidth]{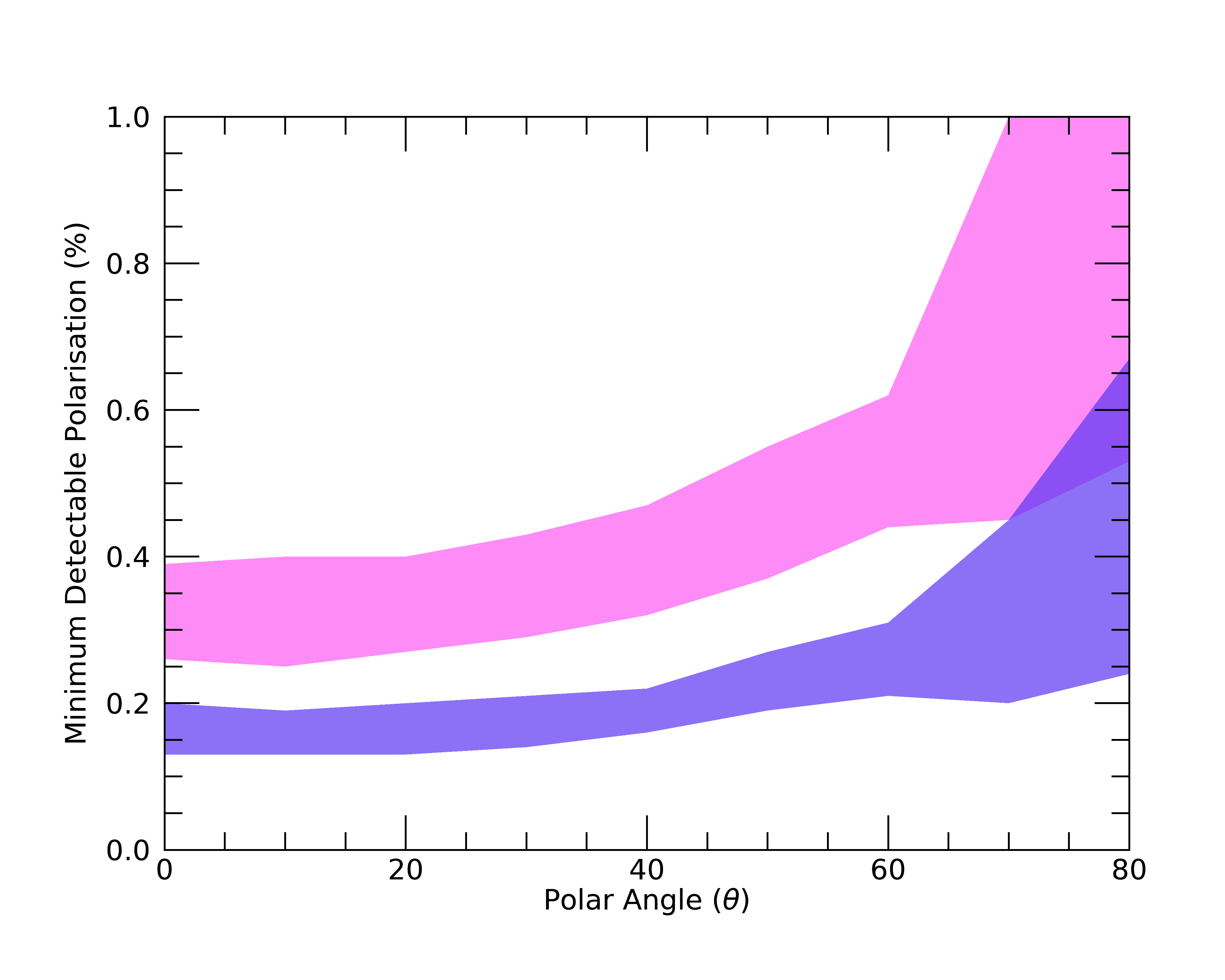}
\end{center}
  \caption
 {\label{fig:MDP} Minimum detectable polarization obtained for incident angles
 $\theta$ = 0$^\circ$ -- 
 80$^\circ$ for two different cases of Compton events: 3000 (pink) and 10,000 (blue).
 The spread in the MDP values within the pink and blue bands corresponds to different PA values.} 
\end{figure}

For this purpose, we performed a set of Geant4 simulations using a single CZT module illuminated by unpolarized beams, having a flat spectrum in the energy range 100 -- 400 keV. Simulations were carried out for incidence angles $\theta$ ranging 0$^\circ$ to 80$^\circ$ at every 10$^\circ$, each for 10 million incident photons resulting in $\sim$30000 -- 60000 Compton events. Normalized ASADs for each case are obtained from these simulations. 
We generated ${10}^5$ realizations of ASADs for unpolarized photons drawing Poisson random variables from the normalized ASAD scaled to the mean source counts. Considering the typical range of Compton events observed by CZTI for GRBs, we use two values for source counts: a typical bright GRB counts of 3000 and an extreme bright case of 10,000 counts. Background ASADs are also generated and added to the source considering a background rate of 40 counts/s (typical Compton background rate in AstroSat-CZTI) and GRB duration of 30 seconds. The background subtracted simulated ASADs of unpolarized emission are then fitted with library of templates generated for the same incident spectrum to obtain PF and PA measurements, from which MDP is estimated.
Figure~\ref{fig:MDP} shows the estimated MDP as a function of $\theta$ for 3000 events (pink) and 10,000 events (purple). The shaded regions represent spread in the MDP values for different PAs. 
From Figure \ref{fig:MDP} it is evident that, 
MDP increases significantly with $\theta$ above 
60$^\circ$. 
 
Our current results show that pixellated CZT detectors are well suited for polarization measurements up to $\theta$ of 45$^\circ$. We also see that these detectors have reasonable sensitivity up to 60$^\circ$, but there may be additional systematic effects in 45-60$^\circ$ range.
An ideal situation is to use multiple CZT detectors over a spherical or hemispherical arrangement (as in the 
proposed Indian mission Daksha, to detect GRB polarization) so that for a given GRB the relative 
$\theta$ lies between 0$^\circ$ -- 45$^\circ$ at least for a fraction of the detectors.

\section{Summary}
\label{sect:Summary} 

The prompt emission polarization measurements have been carried out
for a number of bright GRBs detected with AstroSat-CZTI. Since 
GRBs occur in random directions, and the polarization analysis involves
significant use of Geant4 simulations, it is essential to have experimental 
verification of the off-axis polarization measurement capability of the 
pixellated CZT detectors as well as validation of the Geant4 simulations. 
In this context, we carried out experiments
to estimate the polarization of the partially polarized X-rays generated using laboratory radioactive source. 
We also carried out Geant4 simulations of the actual experimental set-up, including the generation of the partially polarized beam by
$\sim90^{o}$ scattering of X-ray photons. 
Here we summarize the main results from the experiment and its implications $-$
\begin{enumerate}
    \item The measurements of PF and PA obtained from the experiments match closely with those obtained through the simulations. This validates the Geant4 simulation setup that we use for off-axis GRB polarimetry analysis for CZTI. 
    \item We also verified that both the modulation curve fitting method and template fitting method yield similar results, particularly when the number of Compton events is limited as in case of CZTI GRBs. This further validates the methods and results reported in Chattopadhyay et al. (2019)\cite{chattopadhyay19} and Chattopadhyay et al. (2022)\cite{chattopadhyay22} for GRB prompt emission.
    \item The present results demonstrate that the pixellated CZT detectors similar to those used in AstroSat-CZTI can certainly be used for off-axis Compton polarimetry up to the incidence angle of $\sim$60$^{\circ}$. However, at incidence angles between 45-60$^{\circ}$, there might be some systematic effects which  needs to be taken into account while interpreting the measured PF. 
    It would be possible to improve the results for angles $\sim$60$^\circ$ with further optimization in the analysis and modelling. However, in practical case of CZTI GRBs with limited number of Compton events, the systematic effects are expected to be less significant compared to the statistical uncertainties, allowing CZTI to study GRB prompt emission up to angles $\sim$60$^\circ$.
    At incidence angle $>$60$^\circ$, the CZTI type pixellated detectors do not have high enough sensitivity for GRB polarimetry, 
\end{enumerate}

These results will be useful for characterization of future polarimetry instruments that are based on CZT detector array. For example, the same CZT detectors, used in CZTI and in this experiment, are being considered for a proposed mission called 
Daksha, which is dedicated to GRB and EM-counterparts of the gravitational wave sources. Daksha will have a significantly larger effective area than 
AstroSat-CZTI, and our results indicate highly promising prospects for 
polarization measurements of a large number of GRBs with such a dedicated mission. 
 
\appendix    

\acknowledgments 
 The authors thank Keith Jahoda and an anonymous reviewer for their detailed review and comments that greatly helped in improving the clarity of the paper. The Geant4 simulations in this work were performed using the HPC resources at the Physical Research Laboratory (PRL). Research work at PRL is supported by Department of Space, Govt. of India. We thank A. P. K. Kutty, M. K. Hingar, Milind Patil and Dhiraj K. Dedhia from TIFR for their support in carrying out an earlier trial of the experiments presented in this work.
NN acknowledges the financial support
of ISRO under AstroSat archival data utilization program.
 

\bibliographystyle{spiejour}   

\listoftables

\listoffigures

\begin{table}[H]
\caption{The incident PF$_{inc}$, PA$_{inc}$ along with the fitted PA$_{sim}$, PF$_{sim}$ and PA$_{expt}$, 
PF$_{expt}$ corresponding to template fitting and modulation curve fitting method for different incident $\theta$, $\phi$ and PA values. The 1$\sigma$ error bars for respective results are also given. Note: The subscript sim corresponds to simulations results and expt corresponds to experimental results.}
\begin{flushleft}
\centering 
\label{tab:1}
\scalebox{0.7}{
\begin{tabular}{|c|c|c|c|c|c|c|c|c|c|c|c|c|}

\hline
\rule[-1ex]{0pt}{5ex} \multirow{2}{*}{$\theta$ ($^\circ$)} & \multirow{2}{*}{$\phi$ ($^\circ$)}& \multirow{2}{*}{PA($^\circ$)} & \multirow{2}{*}{PF$_{inc}$ (\%)} & \multirow{2}{*}{PA$_{inc}$ ($^\circ$)} & \multicolumn{4}{|c|}{Template Fitting} & \multicolumn{4}{|c|}{Modulation Curve Fitting} \\
\cline{6-13}
\rule[-1ex]{0pt}{3.5ex}   &  &  &   &  & PF$_{sim}$ (\%) & 
PA$_{sim}$ ($^\circ$) & PF$_{expt}$ (\%) & 
PA$_{expt}$ ($^\circ$) & PF$_{sim}$ (\%) & PA$_{sim}$ ($^\circ$) & PF$_{exp}$ (\%) & PA$_{exp}$ ($^\circ$) \\   

\hline

        \rule[-1ex]{0pt}{3.5ex}  30 &   0 &  90 &       61.9 &  89 & $ 68\substack{+ 14 \\ -13}$ & $ 89\substack{+  4 \\  -4}$ & $ 58\substack{+ 12 \\ -12}$ & $ 87\substack{+  4 \\  -5}$ & $      70.1\substack{+  9 \\ -  9}$ & $      89.3\substack{+  2 \\  -2}$ & $      51.5\substack{+  8 \\ -  8}$ & $      86.5\substack{+  3 \\  -3}$ \\
\hline
        \rule[-1ex]{0pt}{3.5ex}  30 &   0 &  45 &       62.5 &  43 & $ 65\substack{+  7 \\  -7}$ & $ 48\substack{+  5 \\  -5}$ & $ 75\substack{+ 10 \\  -9}$ & $ 53\substack{+  4 \\  -4}$ & $      65.8\substack{+ 11 \\ - 11}$ & $      48.2\substack{+  3 \\  -3}$ & $      71.4\substack{+ 14 \\ - 14}$ & $      50.8\substack{+  3 \\  -3}$ \\
\hline
        \rule[-1ex]{0pt}{3.5ex}  30 &  -30 &  90 &       61.4 &  90 & $ 54\substack{+  9 \\ -11}$ & $ 85\substack{+  7 \\  -5}$ & $ 70\substack{+ 11 \\ -10}$ & $ 86\substack{+  5 \\  -5}$ & $      54.1\substack{+ 14 \\ - 14}$ & $      85.2\substack{+  4 \\  -3}$ & $      67.1\substack{+ 15 \\ - 15}$ & $      85.7\substack{+  3 \\  -3}$ \\
\hline
        \rule[-1ex]{0pt}{3.5ex}  30 &  -30 &  45 &       62.3 &  44 & $ 61\substack{+ 11 \\ -11}$ & $ 47\substack{+  3 \\  -4}$ & $ 63\substack{+ 11 \\ -11}$ & $ 47\substack{+  3 \\  -4}$ & $      59.7\substack{+ 11 \\ - 11}$ & $      46.6\substack{+  2 \\  -3}$ & $      61.0\substack{+ 12 \\ - 12}$ & $      45.2\substack{+  2 \\  -3}$ \\
\hline
        \rule[-1ex]{0pt}{3.5ex}  45 &   0 &  90 &       60.8 &  89 & $ 50\substack{+ 17 \\ -16}$ & $ 87\substack{+  7 \\  -7}$ & $ 73\substack{+ 15 \\ -14}$ & $ 92\substack{+  4 \\  -5}$ & $      53.4\substack{+ 11 \\ - 11}$ & $      88.4\substack{+  4 \\  -5}$ & $      62.7\substack{+ 10 \\ - 10}$ & $      92.3\substack{+  3 \\  -3}$ \\
\hline
        \rule[-1ex]{0pt}{3.5ex}  45 &   0 &  45 &       62.9 &  43 & $ 71\substack{+ 12 \\ -11}$ & $ 48\substack{+  7 \\  -6}$ & $ 95\substack{+  5 \\ -15}$ & $ 60\substack{+  4 \\  -5}$ & $      71.9\substack{+ 19 \\ - 19}$ & $      46.2\substack{+  4 \\  -4}$ & $      87.2\substack{+ 22 \\ - 22}$ & $      54.5\substack{+  3 \\  -4}$ \\
\hline
        \rule[-1ex]{0pt}{3.5ex}  45 &  -30 &  90 &       61.9 &  90 & $ 63\substack{+ 10 \\  -9}$ & $ 87\substack{+  5 \\  -5}$ & $ 57\substack{+ 12 \\ -12}$ & $ 85\substack{+  5 \\  -6}$ & $      65.9\substack{+ 12 \\ - 12}$ & $      84.4\substack{+  3 \\  -3}$ & $      50.6\substack{+ 18 \\ - 18}$ & $      83.7\substack{+  6 \\  -4}$ \\
\hline
        \rule[-1ex]{0pt}{3.5ex}  45 &  -30 &  45 &       63.2 &  44 & $ 68\substack{+ 13 \\ -13}$ & $ 46\substack{+  4 \\  -5}$ & $ 79\substack{+ 17 \\ -16}$ & $ 52\substack{+  5 \\  -3}$ & $      69.2\substack{+ 13 \\ - 13}$ & $      45.5\substack{+  2 \\  -3}$ & $      74.0\substack{+ 16 \\ - 16}$ & $      50.4\substack{+  3 \\  -4}$ \\
\hline
        \rule[-1ex]{0pt}{3.5ex}  50 &   0 &  90 &       61.5 &  89 & $ 60\substack{+ 15 \\ -16}$ & $ 88\substack{+  6 \\  -6}$ & $100\substack{+  0 \\ -14}$ & $ 91\substack{+  4 \\  -5}$ & $      60.3\substack{+ 11 \\ - 11}$ & $      87.6\substack{+  3 \\  -4}$ & $      95.7\substack{+ 13 \\ - 13}$ & $      91.8\substack{+  3 \\  -3}$ \\
\hline
        \rule[-1ex]{0pt}{3.5ex}  50 &  -30 &  90 &       59.7 &  91 & $ 55\substack{+ 12 \\ -12}$ & $ 92\substack{+  9 \\  -7}$ & $ 64\substack{+ 17 \\ -17}$ & $ 77\substack{+  8 \\  -8}$ & $      57.3\substack{+ 19 \\ - 19}$ & $      89.9\substack{+  6 \\  -5}$ & $      52.2\substack{+ 28 \\ - 28}$ & $      76.5\substack{+  9 \\  -5}$ \\
\hline
        \rule[-1ex]{0pt}{3.5ex}  60 &   0 &  90 &       59.5 &  90 & $ 84\substack{+ 16 \\ -18}$ & $ 87\substack{+  4 \\  -4}$ & $100\substack{+  0 \\ -14}$ & $ 88\substack{+  5 \\  -8}$ & $      81.9\substack{+ 14 \\ - 14}$ & $      87.0\substack{+  3 \\  -3}$ & $      93.9\substack{+ 21 \\ - 21}$ & $      85.2\substack{+  4 \\  -6}$ \\
\hline
        \rule[-1ex]{0pt}{3.5ex}  60 &   0 &  45 &       63.3 &  43 & $ 72\substack{+ 14 \\ -13}$ & $ 47\substack{+  7 \\  -6}$ & $ 99\substack{+  1 \\ -22}$ & $ 78\substack{+  8 \\  -7}$ & $      74.8\substack{+ 21 \\ - 21}$ & $      46.3\substack{+  4 \\  -5}$ & $      88.5\substack{+ 33 \\ - 33}$ & $      73.0\substack{+  4 \\  -6}$ \\
\hline
        \rule[-1ex]{0pt}{3.5ex}  60 &  -30 &  90 &       60.9 &  90 & $ 82\substack{+ 12 \\ -10}$ & $ 85\substack{+  4 \\  -4}$ & $100\substack{+  0 \\ -19}$ & $ 65\substack{+  7 \\  -4}$ & $      83.9\substack{+ 19 \\ - 19}$ & $      84.1\substack{+  4 \\  -3}$ & $     100\substack{+ 0 \\ - 24}$ & $      61.0\substack{+  4 \\  -3}$ \\
\hline
        \rule[-1ex]{0pt}{3.5ex}  60 &  -30 &  45 &       62.9 &  44 & $ 74\substack{+ 16 \\ -16}$ & $ 47\substack{+  4 \\  -4}$ & $100\substack{+  0 \\ -21}$ & $ 51\substack{+  5 \\  -5}$ & $      75.8\substack{+ 16 \\ - 16}$ & $      46.7\substack{+  2 \\  -3}$ & $     100\substack{+ 0 \\ - 25}$ & $      47.2\substack{+  3 \\  -3}$ \\
\hline

\end{tabular}  
}
\end{flushleft} 
\end{table} 

\end{spacing}

\end{document}